\documentclass[
superscriptaddress,
reprint,
 amsmath,amssymb,
 aps,
]{revtex4-2}

\usepackage{graphicx}
\usepackage{dcolumn}
\usepackage{bm}
\usepackage{hyperref}
\usepackage[mathlines]{lineno}
\usepackage{xcolor}
\usepackage{siunitx}
\DeclareSIUnit\gcc{\gram\per\cm\cubed}
\usepackage{fontawesome5} 

\begin{document}
\title{Hydrodynamic simulations of expanded warm dense foil heated by pulsed-power}

\author{Luc Revello}
\email{luc.revello@cea.fr}
\affiliation{CEA, DAM, DIF, F-91297 Arpajon, France}
\affiliation{Université Paris-Saclay, CEA, LMCE, F-91680 Bruyères-le-Châtel, France}

\author{Laurent Videau}
\affiliation{CEA, DAM, DIF, F-91297 Arpajon, France}
\affiliation{Université Paris-Saclay, CEA, LMCE, F-91680 Bruyères-le-Châtel, France}

\author{Frédéric Zucchini}
\affiliation{CEA, DAM, Gramat, F-46500 Gramat, France}

\author{Mathurin Lagrée}
\affiliation{CEA, DAM, DIF, F-91297 Arpajon, France}
\affiliation{Université Paris-Saclay, CEA, LMCE, F-91680 Bruyères-le-Châtel, France}

\author{Christophe Blancard}
\affiliation{CEA, DAM, DIF, F-91297 Arpajon, France}
\affiliation{Université Paris-Saclay, CEA, LMCE, F-91680 Bruyères-le-Châtel, France}

\author{Benjamin Jodar}
\affiliation{CEA, DAM, DIF, F-91297 Arpajon, France}
\affiliation{Université Paris-Saclay, CEA, LMCE, F-91680 Bruyères-le-Châtel, France}

\date{\today}

\begin{abstract}

Warm Dense Matter lies at the frontier between condensed matter and plasma, and plays a central role in various fields ranging from planetary science to inertial confinement fusion. Improving our understanding of this regime requires experimental data that can be directly compared with theoretical and numerical models over a broad range of conditions. In this work, a pulsed-power experiment is described in which thin metallic foils, confined within a sapphire cell, are Joule-heated to achieve the expanded warm dense matter regime. 
Designing such an experiment is challenging, as it requires simultaneously predicting the electrical response of the pulsed-power driver and the hydrodynamic evolution of the heated material.
To tackle this challenge, a modeling framework has been developed that couples an electrical description of the pulsed-power system, including the driver, the switching stages and the load with a one-dimensional hydrodynamic code.
This coupling allows the electrical energy deposition and the load thermodynamic evolution to be consistently linked through the material electrical conductivity. This approach takes advantage of the simplicity of a 1D geometry while retaining the essential physics and allowing to reproduce various measurements with good accuracy, such as expansion velocity, current and voltage. This numerical approach therefore constitutes a robust and efficient method for designing and optimizing future Warm Dense Matter experiments using pulsed-power facilities.

\end{abstract}


\maketitle


\section{Introduction\protect}
\label{sec:intro}

Understanding the behavior of matter under extreme thermodynamic conditions is a central issue in several domains of physics, including astrophysics, inertial confinement fusion experiments and high-energy-density (HED) science. 

Within this broad landscape of extreme states, a particularly complex subset is the intermediate regime bridging condensed matter and plasma, known as Warm Dense Matter (WDM) \cite{WDM_roadmap}.
This regime occupies a region of the density, temperature domain ($\rho,T$) typically spanning densities from $0.1$ to $10$ times the nominal density, and temperatures between $0.1$ and $10~\mathrm{eV}$. For example, accurate simulation of the onset and growth of magneto-hydrodynamic (MHD) instabilities in HED experiments relies on energy deposition processes and transport properties, which in turn require reliable data in the WDM regime \cite{Stanek_propagate_uncertainties, ICF_Instabilities_2025}. Similarly, in astrophysics, accurate equations of state (EOS) and transport properties are required over a wide range of thermodynamic conditions to describe astronomical objects \cite{howard2023jupiter, PRELIMINARY_JUPITER_MODEL, koester1990physicswhitedwarf}.

This regime is particularly challenging to model because it involves ionic correlation effects between partially ionized atoms as well as electron degeneracy effects \cite{dornheim2023electronic}, that neither classical plasma physics nor conventional condensed matter theories describe correctly.  Quantum molecular dynamics approaches appear to be well suited to describe such media, but they face significant numerical difficulties for systems at intermediate to high temperatures \cite{WDM_roadmap}.

Experimentally, significant effort has been devoted to developing a variety of techniques to produce and diagnose warm matter at nominal and above densities, using high-power lasers \cite{High_laser, High_laser2}, flyer-plate impacts \cite{flyer_laser, flyer_magnetized},  heavy particle-beam heating \cite{Isochoric_1, Isochoric_2, Isochoric_3} and electron beam \cite{electron_beam}.
However, data obtained in the expanded (lower than nominal density) regime is particularly valuable for the construction and validation of EOS and conductivity models relevant to inertial confinement fusion, especially in indirect-drive hohlraum configurations \cite{hohlraum_chen2024}. For exploring this less studied regime, where matter expands from the nominal toward a lower density plasma state, Joule-heating techniques have emerged as efficient platforms \cite{Korobenko_1999,Renaudin_EPI, Korobenko_2005}. In this kind of experiment, the transition from the highly conductive solid phase to melting, vaporization, and ultimately to partially ionized plasma occurs on a sub-microsecond timescale.


To accurately understand and interpret the dynamics of such transitions, numerical modeling is required, typically based on hydrodynamic or magneto-hydrodynamic simulations. In both cases, hydrodynamics plays a key role in the thermodynamic evolution. Depending on the load geometry and current configuration,  a multidimensional resolution can be necessary. This resolution need to be coupled with an external electrical circuit model to account for the time-dependent current delivered to the load \cite{Luo_2012, Chung_2016}.
When the confinement medium exhibits a low acoustic impedance, strong asymmetries can develop during the expansion, leading to significant spatial distribution gradients of density and temperature \cite{filamentation}. These gradients in turn, induce local variations in electrical conductivity, which may trigger magneto-hydrodynamic instabilities \cite{Vlastos_1973} and current filamentation \cite{Wang_2020}, thus requiring a fully coupled MHD treatment \cite{Oreshkin_2020}. Such simulations are computationally demanding, except when the experimental configuration is planar with a high impedance confinement medium, causing the loading expansion to be in a single preferential direction.
Under these conditions, the magnetic field can be neglected, the thermodynamic gradients are kept one-dimensional, allowing the problem to be treated using a 1D hydrodynamic description. This simplification considerably reduces the computational cost while preserving the relevant physics of energy deposition and material response.

In the present study, we focus on a 1D planar configuration. We introduce a numerical approach for modeling any pulsed-power driver as an equivalent electrical circuit, covering generators delivering pulsed currents ranging from hundred of kiloamperes up to several megaamperes. This electrical model is subsequently coupled with the one-dimensional hydrodynamic code ESTHER \cite{ESTHER_1, ESTHER_2, ESTHER_3, ESTHER_4}, developed at CEA-DAM.

This paper is organized as follows, Sec.\ref{sec:exp_protocol} describes the experimental setup considered for the numerical study.  Sec.\ref{sec:Discharge_circuit} presents the electrical modeling of the pulsed-power generator and its validation against short-circuit measurements performed over a wide range of operating conditions. Finally, Sec.\ref{sec:hydro} deals with the coupled simulation framework and details the three approaches used to model energy deposition.

\section{Experimental setup\protect}
\label{sec:exp_protocol}
\begin{figure}[h]
    \centering
    \includegraphics[width=8.5cm]{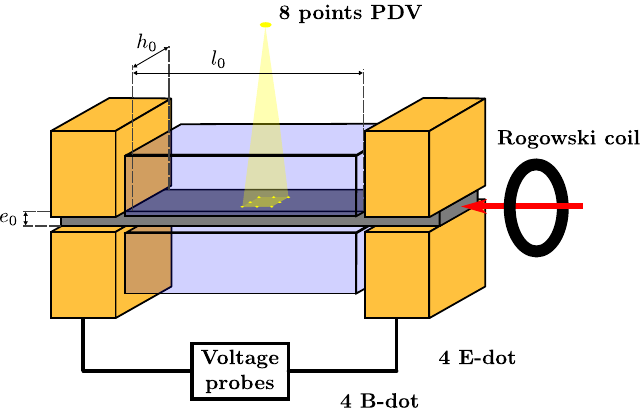}
    \caption{Schematic of the confined exploding foil experiment.}
    \label{fig:exp_explanation}
\end{figure}

This work is mainly motivated by the need to fully simulate an experiment in which a short-duration current pulse from a pulsed-power generator is delivered to a conductive foil load confined within a sapphire anvil. This experiment can be carried out on two main generators: EPP1, which delivers a maximum current of  $140 ~\rm{kA}$ with a rise time of $1.5 ~\rm{\mu s}$, and EPP2, which reaches up to $530 ~\rm{kA}$ with a rise time of $0.83 ~\rm{\mu s}$. In both generators, the load system will exhibit homogeneous conditions and undergo a one-dimensional expansion up to a given time. This allows several quantities to be extracted during the foil thermodynamic evolution.
To place this effort in context, we now provide a detailed description of this experimental configuration (see Fig.\ref{fig:exp_explanation}).

The studied foil is sandwiched between two sapphire plates forming the anvil, each with lateral dimensions of $1 \times 1~\text{cm}^2$ and a thickness between 1 and 5 mm. The foil is characterized by three dimensions: its length $l_0 \approx 10~\mathrm{mm}$, its height $h_0 \approx 6~\mathrm{mm}$, and its thickness $e_0$ in the range of $5$–$20~\mathrm{\mu m}$. The load is electrically connected to the pulsed-power driver along its longitudinal axis through brass clamps. This setup allows a  sub-microsecond controlled transition from a highly conductive solid phase to melting, vaporization, and ultimately partially ionized plasma. These evolutions arise from the combined effects of temperature increases due to resistive (Joule) heating and pressure elevation caused by the foil thermal expansion, as well as the acoustic impedance mismatch between the foil and the confining medium.

The experiment integrates electrical diagnostics, including a Rogowski coil for current measurement and voltage probes as primary diagnostics, the EPP2 is also complemented by four B-dot and four D-dot probes \cite{B_D_dot_wagoner2008}.

An optical diagnostic consisting of an eight-probe Photon Doppler Velocimetry (PDV) system is also implemented on EPP2 to record the expansion velocity in various positions. This diagnostic reduces uncertainty by providing redundancy and allows spatial uniformity to be controlled during the heating process. Whereas EPP1 features a single PDV point but includes a $532 ~\rm{nm}$ laser, the addition of a ruby plate to the target assembly taking advantage of its mechanical equivalence to sapphire and allows this laser beam to excite ruby luminescence. This enables pressure measurements using the well-established Photoluminescence Ruby (PRL) diagnostic \cite{PRL_1972Forman}. The main advantage of using both EPP1 and EPP2 lies in the differences in their current-pulse duration and intensity, which directly determine the thermodynamic trajectory that can be achieved by the target. 

In this way, time-resolved density, internal energy, pressure, and electrical conductivity can then be deduced from these measurements, as detailed in \cite{2024Jodar}.

Although there are significant differences between the two installations, they both rely on a pulsed-power system, in which intense current pulses provide the energy required for a rapid heating of the target.



\section{Discharge circuit\protect} 
\label{sec:Discharge_circuit}
Pulsed-power systems are designed to store a specific amount of energy and release it rapidly within a short-time interval. The stored energy depends quadratically on the charging voltage. Pulsed-power system can have a complex design, but it can  be accurately characterized by three macroscopic parameters that encompass this complexity. 

These parameters are the resistance $R$ defined by Ohm’s law, the capacitance $C$ opposing changes in the time derivative of the voltage, and the inductance $L$ opposing changes in the time derivative of the current.
These three parameters can be used to evaluate the time evolution of the current $i$  flowing through a RLC series circuit, according to the following equation :
\begin{equation}\label{equa:RLC}
    \frac{i}{C} + \frac{d}{dt} \left( L\frac{di}{dt} \right)+\frac{d}{dt} \left( R~i \right) = 0~.
\end{equation}

This equation can be solved analytically for constant parameters $R$, $L$ and $C$. However, in exploding wire experiments, the electrical properties of the wire vary significantly over time. Therefore, it is necessary to numerically solve the current differential equation with time-dependent resistance and inductance, whereas the capacity remains constant.

\subsection{Current solver}
\label{subsec:current_solver}

The current $i$ in Eq.\ref{equa:RLC} can be  discretized as $I$, using a second-order centered finite difference scheme in time, yielding an explicit recurrence relation for the current, such as :

\begin{align} \label{equa:RLC_discretisation}
I^{n+1} &= I^n \left( \frac{- \Delta t^2}{L^n C}+ \Delta t \frac{R^{n-1}-2R^n}{L^n}+\frac{L^n+L^{n-1}}{L^n} \right) \\
& + I^{ n-1} \left( \frac{-L^{n-1}}{L^n}+ \Delta t \frac{R^n}{L^n} \right) ~.
\end{align}

Here, the superscript $n$ denotes the evaluation at the current time step $t^n = n \Delta t$, with $\Delta t$ the time step. The terms involving $\frac{\Delta t^2}{L^n C}$, $\frac{L^n + L^{n-1}}{L^n}$, and $\frac{L^{n-1}}{L^n}$ originate from the discretization of the inductive term $\frac{d}{dt} \left( L \frac{di}{dt} \right)$, while the coefficients containing $R^n$ and $R^{n-1}$ arise from the time derivative of the resistive contribution $\frac{d}{dt}(R\,i)$.  This method offers both accuracy and computational efficiency, while accommodating time-dependent resistance and inductance. 

\begin{figure}
    \centering
    \includegraphics{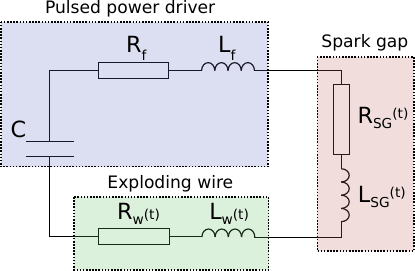}
    \caption{Equivalent electrical circuit used to model a generic pulsed-power system, including the driver stage, the switch and the exploding-wire load.}
    \label{fig:Schema_Elec}
\end{figure}

According to an idealized discharge circuit composed of a pulsed-power driver, a spark-gap switch, and a load (see Fig.\ref{fig:Schema_Elec}), the total time-dependent resistance and inductance can be expressed as:

\begin{align} \label{equa:R_L_sum}
    R(t) &= R_{f} + R_{SG}(t) + R_{W}(t) \\
    L(t) &= L_{f} + L_{SG}(t) + L_{W}(t). 
\end{align}

The first section has constant electrical properties $C$, $R_f$ and $L_f$, reflecting the intrinsic characteristics of the driver. The second section refers to the spark-gap switch which exhibits highly nonlinear behavior.  Indeed, the switch resistance $R_{SG}$ and inductance $L_{SG}$ are initially very high at the moment of switching, then rapidly decrease to much lower values. The third section corresponds to the exploding-wire itself, which undergoes state transitions from solid to plasma induced by resistive heating, thereby altering its electrical properties $R_{W}$ and $L_{W}$ over time.

\subsection{Switch modeling}
\label{subsec:switch_model}

To minimize the influence of the switch response on the current pulse and so the characteristic discharge time, it is essential to use a switch that closes the circuit as quickly as possible. Such ultra-fast switches are known as spark-gaps \cite{SG_low_pressure}. It consists of at least two electrodes separated by an insulating medium, which can be solid (dielectric or semiconductor), liquid, or gaseous (under pressure or in a primary vacuum) \cite{generalite_SG}. The transition from the insulating state to the conducting state is triggered by an external control signal, which can be a laser pulse \cite{laser_SG_switch} or the action of a third electrode positioned near the gap \cite{third_SG_switch} to locally enhance the electric field. In the laser-triggering method, a laser pulse is focused onto the cathode, extracting electrons that are then accelerated by the inter-electrode electric field to initiate the discharge. In the other method, the third electrode increases the local electric field to a level sufficient to initiate breakdown in accordance with Paschen’s law  \cite{Paschen}.
A model developed by Braginskii \cite{braginskii}, is commonly used to describe the behavior of an arc from its initiation to its stable phase, regardless of the triggering method used. This model  describes an arc as a cylinder of constant electrical conductivity $\sigma$ and a time-dependent radius $a (t)$ which can be evaluated from the following equation :

\begin{equation}\label{a_Brag}
    a^2 (t) = \left( \frac{4}{\pi^2 \rho \xi \sigma}\right)^{1/3} \int_0^{t} i^{2/3} (t')~dt' ~.
\end{equation}

This electric arc is characterized by a current $i$, propagating in a gas medium of density $\rho$. The dimensionless coefficient $\xi$ is specific to the atomic composition of the gas (equal to 4.5 for hydrogen and air).

The electrical resistance of a cylinder $R_c$, assuming constant conductivity, is given by:

\begin{equation} \label{Pouillet_cyl}
    R_{c} = \frac{d}{\pi a^2 \sigma} ~,
\end{equation}

where $d$ is the cylinder length, while its inductance $L_c$ is expressed as:

\begin{equation} \label{Inductance_cc}
    L_{c} = \frac{\mu_0 \mu_r}{2 \pi} d \left( \ln \left( \frac{2 d}{a}\right) - \frac{3}{4} \right) ~,
\end{equation}

with $\mu_0$ being the permeability of free space and $\mu_r$ the relative permeability of the conductor. By substituting the constant radius  $a$ in Eqs.\ref{Pouillet_cyl} and \ref{Inductance_cc}, with the time-dependent radius $a(t)$ given by Eq.\ref{a_Brag}, the temporal evolution of the spark-gap electrical properties, namely the resistance $R_{SG}$ and inductance $L_{SG}$  can be determined.

In order to reduce the impact of the switch on the overall system, particularly its resistance and inductance, several identical spark-gaps can be connected in parallel \cite{multi_SP}. Assuming they all trigger simultaneously, the current flowing through each channel will be divided by $N$ the number of channels, the total resistance and inductance of every canals will also be divided by $N$. We then obtain $R_{SG}(t)$ and $L_{SG}(t)$ as follow : 

\begin{equation} \label{R_brag}
    R_{SG} (t) = \left( \frac{\xi \rho}{4 \pi \sigma^2 N} \right)^{1/3} ~\frac{d}{\left(\beta + \int_0^{t} i^{2/3}(t') ~dt'  \right)}~;
\end{equation}

\begin{align}\label{L_brag}
L_{SG} (t) & =  \frac{\mu_0 \mu_r}{2\pi N}\, 
    d \Bigg[\ln(2d) - \frac{3}{4} \nonumber\\[4pt]
    &- \frac{1}{2}\ln\!\left(
        \left( \frac{4}{\pi^2 \rho \xi \sigma N^2} \right)^{1/3}
        \left( \beta + \int_0^{t} i^{2/3}(t')~ dt' \right)
    \right)
\Bigg]~.
\end{align}
A  $\beta$  parameter has been introduced in the current integration to prevent divergence at the initial time steps by limiting the resistance and inductance values to remain high but finite.
\begin{figure*}
    \includegraphics{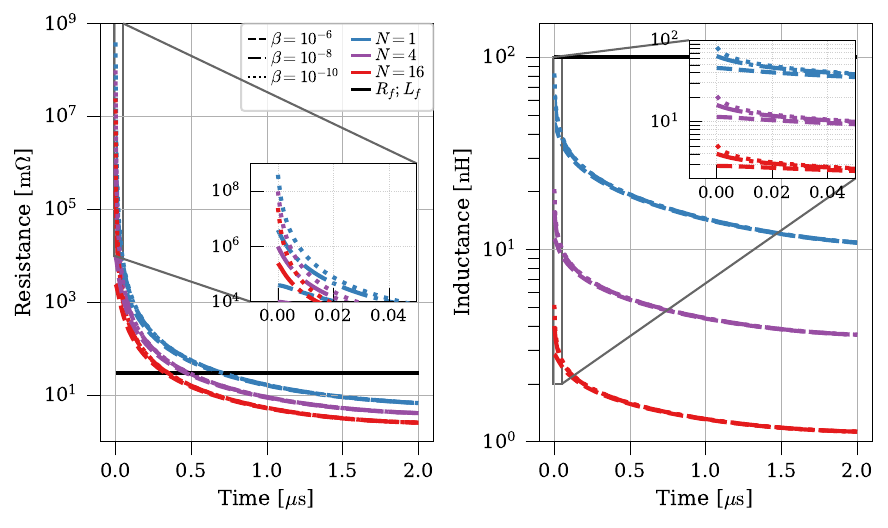}
    \caption{Influence of $\beta$ and $N$ on the resistance and inductance of a generic spark-gap in a generic pulsed-power system (for, $C = 4 ~\rm{\mu F}$, $R_f = 30 ~\rm{m \Omega}$, $L_f = 100 ~\rm{nH}$, charged at $U_0 = 50~\rm{kV}$, with an inter electrode gap $d = 4~\rm{cm}$). Where the line style indicates the value of $\beta$, dotted for $\beta = 10^{-10}$,  dashed-dotted for $\beta = 10^{-8}$ and dashed for $\beta = 10^{-6}$, while the line color denotes the number of channels, blue for $N = 1$, purple for $N=4$ and red for $N=16$.}
    \label{fig:Impact_ecl}
\end{figure*}

For a generic spark-gap in a generic pulsed-power system, Fig.\ref{fig:Impact_ecl} shows that, during the early phase of the discharge ($t < 0.5~\rm{\mu s}$), the spark-gap resistance displays a non-negligible contribution relative to the fixed circuit components $R_f$ and $L_f$.
The spark-gap resistance initially exceeds the fixed resistance by more than seven orders of magnitude, before decreasing to about one third of it. Similarly, the spark-gap inductance starts at roughly $80 \%$ of the fixed inductance and then drops slightly above $1\%$ of it.
Beyond $0.5~ \rm{\mu s}$, both quantities asymptotically approach nearly constant values, still non-negligible, but significantly smaller than their early-time peaks. 

It also shows that $\beta$ primarily affects $R_{SG}$ and $L_{SG}$ at the initial phase but has negligible impact thereafter. In contrast, the parameter $N$ has relatively little effect on the resistance at early times but becomes significant at later stages. Additionally, $N$ has a substantial influence on the inductance throughout the time interval considered. Based on this modified version of the Braginskii model, the spark-gap resistance and inductance are computed at each time step, and these updated values directly feed back into the current calculation for the following step.

\subsection{Short Circuit Validation}
\label{subsec:short_circuit}


To validate our modeling approach, we compare the predicted current waveforms with short-circuit measurements. In addition to the two generators used for the confined expanding-foil experiments, two other pulsed-power generators (located at CEA Gramat) are included in this comparison. First one, the EOLE facility wich is specifically designed to investigate high-energy explosions using scaled-down experiments. It employs a pulsed-power driver capable of delivering a high-current pulse with a peak amplitude of up to $640~\rm{kA}$  and a rise time of $0.9~\rm{\mu s}$. The current is injected into a planar wire, generating a spherical-like  blast wave in the surrounding air, which subsequently interacts with a test model. The second is the GEPI2 platform, wich can deliver currents up to  $4.93 ~\rm{MA}$ with a rise time of $1.23~ \rm{\mu s}$. It is designed to generate high magnetic pressure within a strip line where the samples are placed, targeting a pressure range from $0.1~ \rm{GPa}$ to $50~ \rm{GPa}$. By including these two additional generators in the comparison, we assess the robustness of our simulation framework across generators producing pulsed currents ranging from hundred of kiloamperes to several megaamperes.

In short-circuit configuration, only the resistance $R_f$ and inductance $L_f$ of the pulsed-power driver together with those of the spark-gap, $R_{SG}$ and $L_{SG}$, are taken into account, thereby eliminating all complications related to the target dynamics.
In practice, regardless of  the pulsed-power generator considered, direct experimental determination of their  $R_f$ and $L_f$ values is highly challenging. These quantities can instead be estimated through electromagnetic simulations. In our case, we determined them using current signals with a multi-step optimization based on the least squares method. 
We first fitted  $R_f$ and $L_f$ using the later part of the current pulse, where the overall circuit resistance and inductance can be assumed constant.
Using the previously determined  $R_f$ and $L_f$, we then optimized the parameter $\beta$ over the early-time region, where the spark-gap shows a strong influence.
A final refinement of $R_f$ and $L_f$ was performed in order to subtract the constant yet non-negligible late-time contribution of the spark-gap resistance and inductance.

Also, multi-gap switches are employed in the EPP2, EOLE, and GEPI2 facilities. The actual number of channels $N$ initiating the breakdown is voltage-dependent: higher inter-electrode voltages correspond to an increased number of channels triggering simultaneously. This relationship is empirically established through experiments performed at various charging voltages. All parameters used for simulating the short-circuit conditions of all considered facilities are given in the Table.\ref{tab:facility_specs}.

\begin{table}[h!] 
\centering
\begin{tabular}{|l|c|c|c|c|}\hline
 & EPP1 & EPP2 & EOLE & GEPI2 \\
\hline
$C (\rm{\mu F})$  & $3.96$ & $4.1$ & $5.65$ & $23.8$ \\
$R (\rm{m\Omega})$ & $23$ & $22$ & $10$ & $6.9$\\
$L (\rm{nH})$ & $230$ & $85$ & $46$ & $24.8$ \\
$U_{\text{load}} (\rm{kV})$  & $[25-40]$ & $[35-75]$ & $[30-75]$ & $200$ \\
$N$ & $1$ & $[8-16]$ & $[16-32]$ & $288$ \\
$d (\rm{cm}) $ & $1.04$ & $5.28$& $5.28 $  & $18.2$ \\
$P_{relativ}$ &$[1-3]$ &$1$ &$1$ & $1$\\
$\beta$ &$10^{-9}$ & $10^{-9}$&$5.10^{-8}$ & $10^{-8}$ \\
$I_{\max} (\rm{kA})$ & $\sim 140$ & $\sim 530$ & $\sim 640$ & $\sim 4930$ \\
$ \tau_{rising}(\rm{\mu s})$ & $\sim 1.5$ & $\sim 0.83$ & $\sim 0.9$ & $\sim 1.23$ \\
\hline
\end{tabular}
\caption{Characteristics of the different facilities considered in this study}
\label{tab:facility_specs}
\end{table}

\begin{figure}
    \includegraphics{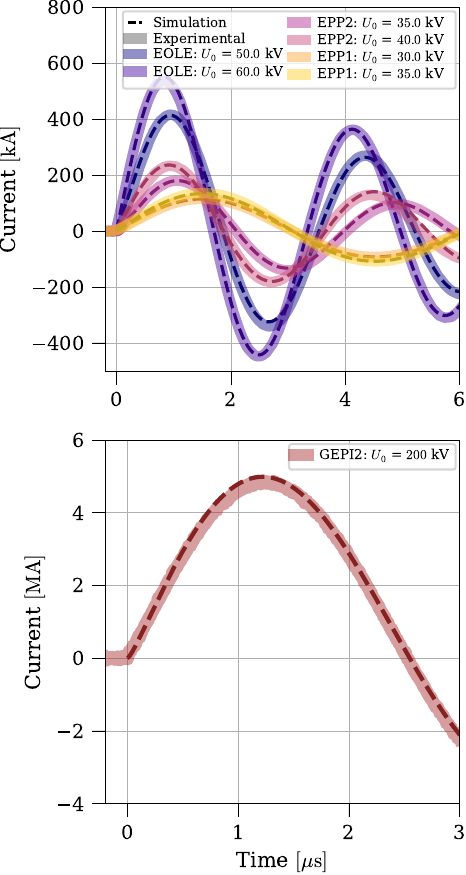}
    \caption{Short-circuit current measurements from four pulsed-power generators, compared with simulations using our discharge model (with parameters in Table.\ref{tab:facility_specs}). Solid lines represent the measured currents, with line thickness indicating the experimental uncertainty, while dashed lines correspond to the simulated currents. Colors identify the different facilities and their respective charging voltages.}
    \label{fig:CC_comparison}
\end{figure}

The Fig.\ref{fig:CC_comparison} shows the resulting currents calculated according to the four generators considered here. These signals are compared with the corresponding experimental measurements for loading voltages ranging from $\SI{35}{\kilo\volt}$ to $\SI{200}{\kilo\volt}$.

Note that the current measurement on GEPI2 after $3 ~\mu s$ is affected by a saturation phenomenon resulting from an experimental limitation. For this reason, only the signal prior to this time is considered.

Fig.\ref{fig:CC_comparison}, shows a very good agreement between experimental and numerical signals over a wide range of discharge currents ranging from $\SI{50}{\kilo\ampere}$ to $\SI{5}{\mega\ampere}$, with rising edges comprised between $0.8-\SI{1.5}{\micro\second}$. 
These results validate the proposed modeling approach, including both the numerical scheme used to solve the current equation and the spark-gap model. They also support the optimization procedure employed to determine the intrinsic circuit properties.

\section{Modeling the heating process\protect}
\label{sec:hydro}

To complete our modeling approach, the next step consists of accurately tackling the modification of the foil thermodynamic and transport properties over the heating process.
In this section, we introduce the hydrodynamic code that will be coupled to the current solver part Sec.\ref{subsec:esther}. Then we present the various validation steps in Sec.\ref{subsec:DM}–\ref{subsec:Sigma_imposed} required to develop a self-consistent numerical framework for coupling hydrodynamics and electrical energy deposition in Sec.\ref{subsec:Coupled_Simulation}.


\subsection{Hydrocode\protect}
\label{subsec:esther}

The ESTHER code \cite{ESTHER_1,ESTHER_2,ESTHER_3,ESTHER_4} is a one-dimensional Lagrangian hydrodynamic solver developed at CEA-DAM, originally designed for simulating laser–matter interactions and shock wave propagation in condensed media. It covers a wide range of timescales, from femtosecond laser–plasma dynamics to the nanosecond regime, where it serves as a reference tool for modeling laser-induced shock phenomena. In the present study, it is applied to a rather different physical system, i.e., the energy deposition originates from Joule heating rather than from a laser beam, and the relevant timescale shifts to the microsecond regime.

The fluid motion is described by the classical conservation equations of mass, momentum, and energy, written in the Lagrangian formalism for each computational cell. In one-dimensional planar geometry, these equations read:
\begin{align}
\frac{D \rho}{D t}& + \rho\frac{\partial u}{\partial r} = 0, \label{eq:mass}\\
\frac{D u}{D t}& + \frac{1}{\rho} \frac{\partial (P - s)}{\partial r} = 0, \label{eq:momentum}\\
\frac{D E_i}{D t}& + \frac{1}{\rho} (P - s) \frac{\partial u}{\partial r} =0. \label{equa:hydro_energy_conservation}
\end{align}

Here, $r$ denotes the cell position, $t$ the time, $u$ the material velocity, $\rho$ the mass density, $P$ the matter pressure, $s$ the deviatoric stress contribution, $E_i$ the specific internal energy and $D$ the material derivative defined as :
\begin{equation}
    \frac{D}{Dt} = \frac{\partial}{\partial t} + u \frac{\partial}{\partial r} ~.
\end{equation}

These equations are solved using a finite volume scheme, ensuring the conservation of mass, momentum, and total energy across each cell interface.\\

To close the hydrodynamic system, ESTHER implements an EOS that relates thermodynamic quantities such as $P(\rho,E_i)$ and $T(\rho,E_i)$. These provide a consistent description of phase transitions and a reasonably accurate description of material response over several orders of magnitude in temperature and density.\\
Thermal conduction is modeled through a diffusive heat flux defined by Fourier’s law,
where the thermal conductivity depends on the local thermodynamic state, using tabulated data from the literature, such as the Y. S. Touloukian tables \cite{Touloukian}, or from average atom calculations \cite{SCAALP_2004, SCAALP_2010}.\\ The code also includes  a radiation energy transfer module based on the discrete ordinates ($S_N$) method \cite{SN_modest2021}. However, under the present conditions, radiative effects were found to be negligible, so radiation transport was omitted from the simulations.\\

The ESTHER code, therefore provides a comprehensive and physically consistent description of energy deposition and thermodynamic evolution in condensed matter under extreme conditions. Its architecture enables a tight coupling between hydrodynamics, equations of state, and energy transport mechanisms, ensuring a description of transient high-pressure and high-temperature phenomena. The validity of such simulations is therefore strongly tied to the accuracy and reliability of  the material databases on which it relies. \\

As reported  in our previous work \cite{2024Jodar}, the magnetic contribution to the metallic foil dynamics can be neglected in the confined pulsed Joule-heating experiment. Consequently, the heating term arising from the Joule effect is directly incorporated into Eq.\ref{equa:hydro_energy_conservation} by replacing the right-hand side zero with a source term $S_{\epsilon}$, representing the volumetric energy deposition.\\

\begin{figure*}
    \centering
    \includegraphics[width=15cm]{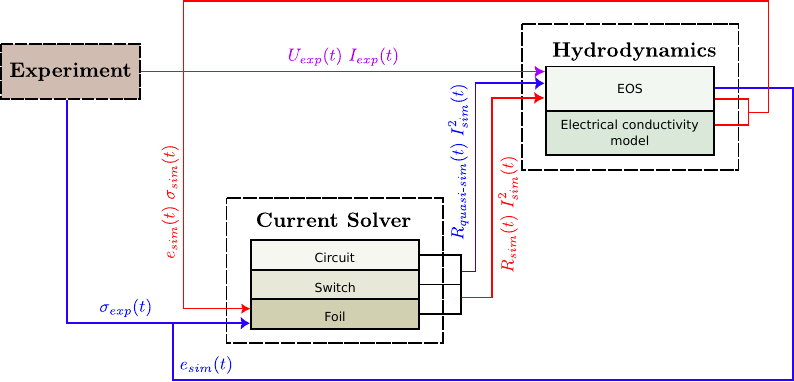}
    \caption{Different coupling paths used for the energy-deposition term in the hydrodynamic simulations. The purple path relies directly on the experimental current and voltage measurements; the blue path uses the experimental electrical conductivity together with the discharge-current model; and the red path is fully numerical, combining the discharge-current model with a tabulated conductivity model.}
    \label{fig:Schema_couplage}
\end{figure*}

In the following, we describe the various schemes (see Fig.\ref{fig:Schema_couplage}
) used for dealing with the Joule heating deposition.  To assess the impact of the deposition scheme on the thermodynamic evolution, we compare simulations with an experiment carried out with an aluminium foil load of a ($1.04~\text{cm}$) length, a height of ($5.9~\text{mm}$), and an initial thickness of ($18.3~\rm{\mu m}$) (see Fig.\ref{fig:exp_explanation}). It is confined between two  ($3~\text{mm}$) thick sapphire plates.
This load was electrically connected to the EPP2 generator charged at 35 kV.\\
The assumption of uniform heating and one-dimensional expansion remains valid up to the end of the expansion velocity signal, i.e., up to $0.47~\rm{\mu s}$. Within this time window, the measured current, voltage, and velocity allow access to several derived quantities, including pressure, internal energy, and electrical conductivity \cite{2024Jodar}.

\subsection{Hydrodynamic validation: experimental power used as source term \protect}
\label{subsec:DM}

As a first step, the deposited power is derived from the experimental current $I_{exp} (t)$ and voltage $U_{exp} (t)$ measurements that originate from the Joule heating process. This power is directly introduced into the code as a source term of power deposition, corresponding to the purple path displayed in Fig.\ref{fig:Schema_couplage}. In this scheme, the source term is applied exclusively and uniformly throughout the entire foil thickness. Then, we compare  the numerical results to the measured expansion velocity, density, and internal-energy variation. Simulations are performed using the following three equations of state : SESAME \cite{SESAME}, BLF \cite{BLF}, and Hébert \textit{et al.} \cite{2023Hebert}.

As demonstrated in a previous study \cite{2024Jodar}, the experimental density profile remains spatially uniform and is calculated from the displacement of the aluminum-sapphire interface $x(t)$ obtained by temporal integration of the velocity signal $x(t)={\displaystyle \int_0^{t}{v}(t)~dt}$, leading to :

\begin{equation}
\begin{split}
\rho (t)& = \frac{{m}_{0}}{{l}_{0}{h}_{0}e(t)} \\
 & = \frac{{m}_{0}}{{l}_{0}{h}_{0}{d}_{0}(1+2x(t)/{e}_{0})},
\end{split}
\label{eq:eos_rho_calc} 
\end{equation}

where ${m}_{0}$ denotes the foil initial mass and $e(t)$ the time-dependent foil thickness. Concerning the internal energy variation $\Delta{E}_{i}(t)$, it is calculated by an energy balance including the electrical energy deposited ${E}_{dep}$, the kinetic energy variation$ \Delta{E}_{k}$ and the mechanical work $W$ as :

\begin{equation}
\begin{split}
\Delta {E}_{i} & = {E}_{dep} - W - \Delta{E}_{k} \\
\Delta E_i (t) &= \int_{0}^t P_{dep}(t') dt' - \int_{0}^t p(t') dV(t') \\
& - \frac{1}{2} m_0U^2_{int}(t) 
\end{split}
\label{eq:calc_Ei} 
\end{equation}  

with $P_{dep}(t)$ the deposited electrical power, $p(t)$ the pressure, $V(t)$ the volume, and $U_{int}(t)$.
\begin{figure*}
  \centering
  \includegraphics{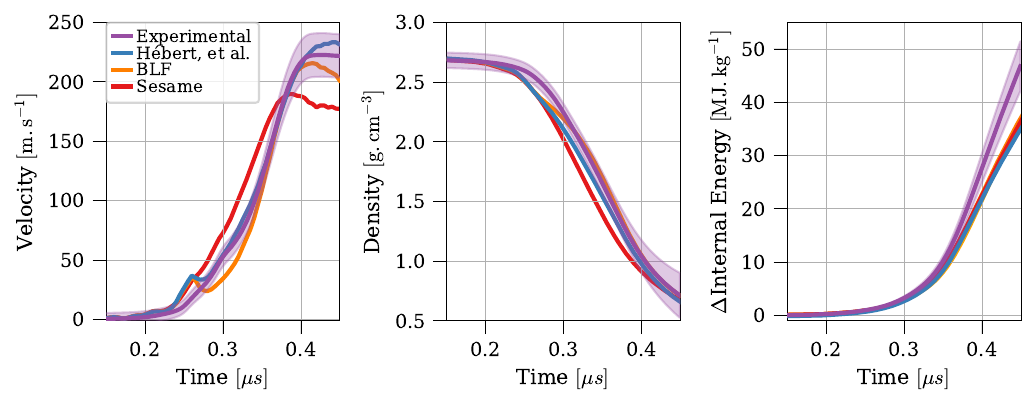}
  \caption{Simulations performed using the experimental power signal as the source term. Experimental expansion velocity (a), density (b), and internal energy (c) are shown in purple and compared with numerical results obtained using the SESAME \cite{SESAME} (red), BLF \cite{BLF} (orange), and Hébert \textit{et al.} \cite{2023Hebert} (blue) equations of state.}
  \label{fig:DM_ei}
\end{figure*}

In Fig.\ref{fig:DM_ei} are displayed the temporal profiles of the interface velocity (Fig.\ref{fig:DM_ei}.(a)), foil density (Fig.\ref{fig:DM_ei}.(b)) and foil internal energy variation (Fig.\ref{fig:DM_ei}.(c)) calculated using the three EOS models considered. When compared to experimental data, a quite good agreement is observed with BLF and Hébert \textit{et al.} EOS up to $\sim \SI{0.25}{\micro\second}$. In particular, predicted velocity signals (Fig.\ref{fig:DM_ei}.(a)) are very close to each other up to $\sim \SI{0.25}{\micro\second}$, where the solid to liquid transition occurs around $\sim \SI{40}{\meter\per\second}$. After $0.25 ~ \rm{\mu s}$, the difference between the three simulations suggests that the modeling of melting is clearly tackled in different ways, causing either a tenuous kink with SESAME , a smooth transition with Hébert \textit{et al.}, or a sharp velocity drop with the BLF model. Experimentally, this kink is found around $\sim \SI{50}{\meter\per\second}$ and does not cause a velocity drop. Beyond this transition, the Hébert \textit{et al.} and BLF models surround the experimental velocity up to $\sim \SI{0.4}{\micro\second}$, contrary to the SESAME model. The rising-edge is better reproduced by the BLF model while the model of Hébert \textit{et al.} is the only one reproducing the maximum velocity.

Regarding the density evolution, good agreement is observed for both BLF and Hébert \textit{et al.} simulations, while SESAME displays a bigger temporal shift originating from the poor velocity prediction. In general, all simulated density profiles begin to decrease slightly earlier than observed experimentally, mainly due to the solid–liquid transition kink, which induces a rapid drop in density.
Following this transition, the BLF prediction is superimposed with the experimental profile. In accordance with the velocity signal, the model of  Hébert \textit{et al.} is located between the two previous models and remains consistent with the experimental data.
Regarding the variation in internal energy, an excellent agreement is observed between all simulations and the experimental data up to approximately $0.35~\rm{\mu s}$. Beyond this value, the experimental curve show a markedly steeper increase than any of the simulations, resulting in an internal energy variation about $20 \%$ higher at $0.47~\rm{\mu s}$. Among the models, the BLF EOS provides the closest prediction, while the Hébert \textit{et al.} internal energy shows a slightly larger deviation.\\

Based on these observations, numerical results highlight the influence of the equation-of-state model used to reproduce our experimental data, notably around the solid to liquid transition. Consequently, the BLF and Hébert \textit{et al.} EOS appear to be more appropriate in this regime than the SESAME one. The overall good agreement observed validates the approach of using the experimental power as a source term for selecting the most adequate EOS model. 

\subsection{Electrical validation: experimental conductivity used as source term \protect}
\label{subsec:Sigma_imposed}

Once the hydrodynamic part validated, we propose to use the experimental conductivity signal $\sigma_{\text{exp}}(t)$ as source term to validate the ESTHER computation of current and voltage. The corresponding calculation scheme is illustrated by the blue path in Fig.\ref{fig:Schema_couplage}. With this second method, we no longer rely on the electrical power calculated from the voltage $U_{exp}(t)$ and current $I_{exp}(t)$ measurements. Instead, the power is calculated with the foil resistance ${R}_{W}$ and the computation of $i(t)$ by using the current solver where the electrical properties obtained in short-circuit validation are implemented.

Here, the experimentally measured time-resolved electrical conductivity, $\sigma_{\text{exp}}(t)$, was applied uniformly across the foil thickness. The instantaneous resistance of the foil is calculated as follow:

\begin{equation} \label{equa:R_W_e_tot}
    R_{W}(t) = \frac{l_0}{\sigma_{exp}(t) e_{sim}(t)h_0}
\end{equation}

where $e_{\text{sim}}$ denotes the total foil thickness computed at each time step. The foil inductance $L_W$ is also calculated by the relation in \cite{Piatek_2012} as  : 

\begin{align}
L_W (t) &= \frac{\mu_0}{6 \pi h_0^2} \Bigg[
3 h_0^2 l_0 \ln\left( \frac{l_0 + \sqrt{l_0^2 + h_0^2}}{h_0} \right) \notag\\
& - \left(l_0^2 + h_0^2\right)^{\frac{3}{2}} + l_0^3 + h_0^3 \notag\\
& + 3 h_0 l_0^2 \ln\left( \frac{h_0 + \sqrt{l_0^2 + h_0^2}}{l_0} \right)\Bigg] \\ 
& - \frac{e_{sim}(t)l_0}{h_0}C_{Hoer} \times 10^{-3},
\end{align}
with $C_{Hoer}$ a constant tabulated  in \cite{L_hoer1965}.
Both $R_W$ and $L_W$ play a predominant role in calculating the current $I_{sim}$. At the beginning of the experiment, the total resistance of the discharge circuit is almost solely defined by the circuit and the switch intrinsic properties. However, during the heating process, the foil resistance can exceed the intrinsic resistance of the circuit, thus driving completely the current flow and so the Joule power deposition as ${P}_{dep,sim}(t) = R_W(t) {{I}_{sim}}^2(t)$. This feedback loop ensures a consistent coupling between the electrical circuit and the hydrodynamic evolution of the material, without depending on an electrical conductivity model. 

\begin{figure*}
    \centering
    \includegraphics{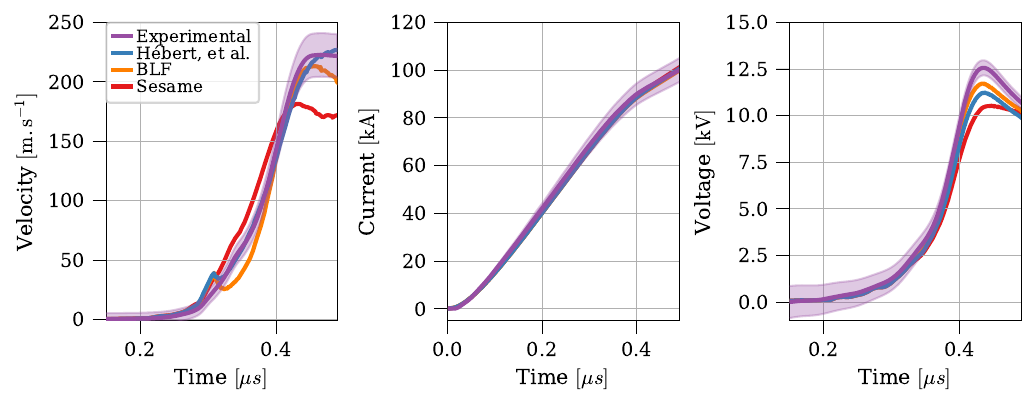}
    \caption{Simulated quantities obtained by imposing the experimental DC electrical conductivity as the source term. Experimental data are shown in purple and compared with the expansion velocity (a), current (b), and voltage (c) calculated using the SESAME \cite{SESAME} (red), BLF \cite{BLF} (orange), and Hébert \textit{et al.} \cite{2023Hebert} (blue) equations of state.}
    \label{fig:sigma_imposed}
\end{figure*}

Fig.\ref{fig:sigma_imposed}.(a), shows similar discrepancies on the velocity as in Fig.\ref{fig:DM_ei}. Although the overall temporal agreement is slightly improved. The solid to liquid transition peak is also too noticeable for both the BLF and Hébert \textit{et al.} models, while it is more tenuous with SESAME. In addition, the velocity plateau predicted by SESAME is once again about $25\%$ lower than the experimental value, while the BLF prediction shows no well-defined plateau after reaching its maximum.

In addition to thermodynamic quantities, this second numerical approach gives access to the electrical behavior. Where Fig.\ref{fig:sigma_imposed}.(b) show the excellent agreement of simulated current regardless of the equation-of-state model. As for the voltage (see Fig.\ref{fig:sigma_imposed}.(c)), the agreement is also very good up to approximately $\sim \SI{0.4}{\micro\second}$, right before the peak voltage. Indeed, the maximum voltage predicted with the three EOS   reaching up to $\sim 10 \: \%$ deviation. This deviation is directly linked to a slight overestimation of the current after $\SI{0.4}{\micro\second}$, which indicates that the simulated foil resistance $R_{\mathrm{sim}}$ drops below the experimental value.
Since the electrical conductivity $\sigma_{\text{exp}}$ is used in both simulations, this discrepancy can only originate from the foil thickness calculation $e_{sim}(t)$, which appears to be slightly underestimated in the simulation.\\

However, since the BLF and Hébert \textit{et al.}  EOS provide the best reproduction of the voltage signal and, consequently, energy deposition, they are retained for the remainder of this study.

\begin{figure}
    \centering
    \includegraphics{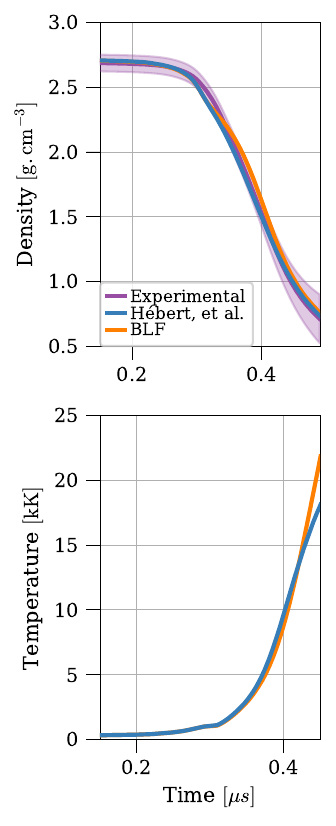}
    \caption{Simulated thermodynamic path obtained by imposing the experimental DC electrical conductivity as the source term. The experimental density is shown in purple and compared with the density calculated using the BLF (orange) and Hébert \textit{et al.} (blue)  EOS . The corresponding temperature obtained from the same simulations is also displayed.}
    \label{fig:rho_T_from_sigma_t}
\end{figure}

With this level of agreement, the simulation can be used to infer additional thermodynamic quantities, in particular the temperature, while the density evolution can be compared to the experimental measurements (see Fig.\ref{fig:rho_T_from_sigma_t}).

Both simulated density profiles show very good agreement with the experiment.
Concerning the temperature, the two simulated profiles show a pretty good agreement up to $15 ~000~\rm{K}$ beyond which a deviation becomes apparent. At $0.45 ~ \rm{\mu s} $, the Hébert \textit{et al.} EOS  predicts a temperature of approximately $18~000 \rm{K}$ , while the BLF EOS  reaches about $21~750 ~ \rm{K}$.

This section allowed us to test the implementation of our current solver within the ESTHER code, showing very good agreement with the experimental current and voltage signals, as well as an expansion velocity similar to the experimental power source term approach. Beyond simply validating our coupling approach, this also provides access to other quantities that cannot be measured directly, such as temperature.

\subsection{Hydrodynamic and electrical coupling\protect}
\label{subsec:Coupled_Simulation}
In order to make the simulations predictive and capable of guiding future experimental campaigns toward specific thermodynamic paths, they must be independent of experimental input data. Unlike the two previous simulation approaches, the power deposition is  handled entirely numerically here through the use of an electrical conductivity model, denoted by the red path in Fig.\ref{fig:Schema_couplage}. This method provides a direct assessment of the relevance and consistency of the database used i.e, (EOS) and electrical conductivity models for the reproduction of the various measured physical quantities.
To compute the electrical conductivity, the code evaluates the local thermodynamic state of each cell, specifically the temperature and density and retrieves the corresponding electrical conductivity from the chosen tabulated model. As a result, if the thermodynamic conditions are not spatially uniform, each cell will receive independent Joule power, according to their temperature and density state. Thus, the macroscopic foil resistance $R_W(t)$ is calculated as the sum of local cell resistances $R_i$, set in parallel, as follow :

\begin{equation}
R_W (t) = \left( \sum_{i=1}^{n} \frac{1}{R_i (t)} \right)^{-1}
\end{equation}

where $R_i(t)$ is given by:

\begin{equation}
    R_i (t)=\frac{\rho_i l_0^2}{m_i\sigma_i(\rho_i, T_i)}
\end{equation}

with $\rho_i$ the cell density, $m_i$ the cell mass, and $\sigma_i(\rho_i, T_i)$ the electrical conductivity obtained from an appropriate conductivity model. This formulation allows for a possible non-uniform electrical conductivity distribution across the foil section, causing a possible non-uniform current density profile. This allows us to verify the limits of the one-dimensional experiment assumption. 

For the sake of simplicity, we present results obtained with a single electrical conductivity model. Because it provides a consistent description over a wide range of thermodynamic conditions, we choose the model proposed by Lee and More modified by Desjarlais \cite{Lee_More_1984, LMD}. 
In the spirit of the empirical coefficient $p2$ introduced by Desjarlais, we modify the original model by introducing a temperature-dependent function $f(T)$ used as a multiplicative factor of the electrical conductivity. This function is defined as :

\begin{equation} \label{equa:f(T)}
f(T)=
\begin{cases}
y_1, & T < T_1, \\[4pt]

\dfrac{y_2-y_1}{T_2-T_1}\,(T-T_1)+y_1,
& T_1 \le T < T_2, \\[6pt]

\exp\!\left[
-\dfrac{1}{x_1}
\left(
\dfrac{T-T_2}{T_2}+x_2
\right)
\right]+k,
& T \ge T_2.
\end{cases}
\end{equation}

Parameters used in Eq.\ref{equa:f(T)} are listed in the Table.\ref{table:param_f_t} they depend on the EOS retained and are optimized in order to match both experimental results published for the solid state \cite{sigma_solid_1} and those recorded in our experiments.
\begin{table} 
\centering
\begin{tabular}{lccccccc}
\hline
EOS & $T_1$ & $T_2$ & $y_1$ & $y_2$ & $x_1$ & $x_2$ & $k$ \\
\hline
BLF & 850 & 4300 & 0.42 & 1.5 & 0.56 & $0.02$ & 0.658 \\
Hébert \textit{et al.} & 850 & 4000 & 0.42 & 1.3 & 0.7 & $0.25$ & 0.67 \\
\hline
\end{tabular}
\caption{Parameters used in  Eq.\ref{equa:f(T)} for the two EOS considered}
\label{table:param_f_t}
\end{table}

 \begin{figure*}
    \centering
    \includegraphics{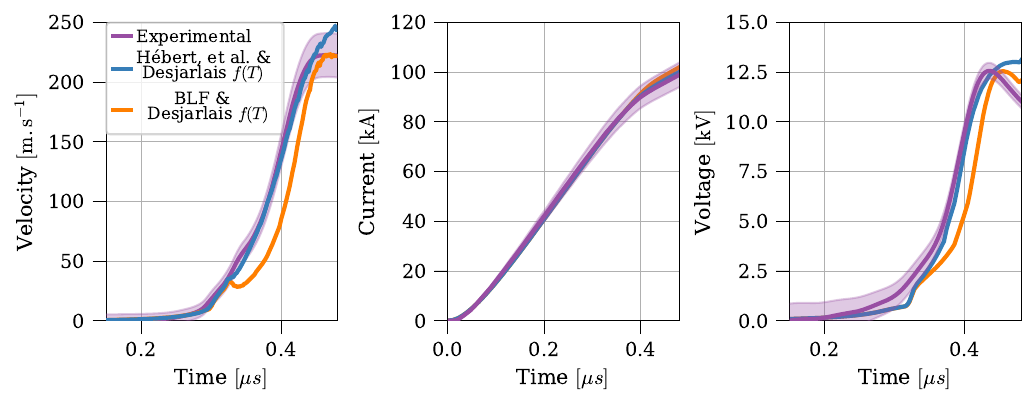}
    \caption{Simulated quantities obtained using the fully simulated scheme. Experimental data are shown in purple and compared with the expansion velocity (a), density (b), and internal energy (c) calculated using the BLF \cite{BLF} and Hébert \textit{et al.} EOS, combined with the modified electrical conductivity model Desjarlais $f(T)$.}
    \label{fig:full_sim_LMD_vs_exp}
\end{figure*}

As shown in Fig.\ref{fig:full_sim_LMD_vs_exp}, the overall trend is well captured by the simulation, particularly the current signal, wich is reproduced very accurately. The onset of motion in the velocity signal is also reasonably well resolved. After approximately $0.33 ~\rm{\mu s}$, a decrease in velocity is observed with the BLF EOS, similarly to the other approaches using the same EOS. In the present case, this decrease induces a temporal shift in the velocity rise, which is not recovered at later times. Nevertheless, the velocity plateau reaches the correct amplitude. The use of Hébert \textit{et al.} EOS shows a good reproduction of the velocity signal up to $0.45~ \rm{\mu s}$. After this time, the experimental plateau is not correctly reproduced. 

An examination of the voltage signals clearly reveals the strong correlation between voltage and expansion velocity. The simulated voltage remains slightly below the experimental curve up to approximately $0.31~\rm{\mu s}$, where a step-like feature appears. 
Beyond this point, the simulated voltage from the BLF EOS follows the experimental trend but exhibits the same temporal shift observed in the velocity signal. The peak amplitude and the subsequent decrease, which are linked to the increase in electrical conductivity, are also well reproduced but are shifted in time. The voltage obtained using the Hébert \textit{et al.} EOS accurately reproduces the rising feature  but does not show the decrease after  $0.45~ \rm{\mu s}$, displaying only the beginning of a plateau.

When comparing the simulated voltage obtained in Fig.\ref{fig:sigma_imposed}  and Fig.\ref{fig:full_sim_LMD_vs_exp}, the approach using an electrical  conductivity model leads to somewhat less accurate agreement with the experimental signal. However, this discrepancy should not be directly attributed to deficiencies in the electrical conductivity model alone.

If we focus on the simulation using BLF EOS, for times prior to $0.33~\rm{\mu s}$, where the velocity in Fig.\ref{fig:full_sim_LMD_vs_exp} is still well reproduced and, consequently, the density remains reliable, both the density and the current are accurately simulated. 
Under these conditions, any subsequent mismatch in energy deposition can only be attributed to the electrical conductivity.  However, the electrical conductivity depends not only on the validity of the conductivity model itself but also on the accuracy of the underlying density and temperature.  While the density is well reproduced, the temperature cannot be accessed experimentally. As a result, it is not possible to unambiguously determine whether the observed discrepancy originates from the equation-of-state or from the electrical conductivity model.

A similar ambiguity is reflected in the velocity signal obtained with the  Hébert \textit{et al.} EOS, where the absence of a clear velocity plateau after $0.45 ~\rm{\mu s}$ suggests a comparable interplay between EOS-dependent temperature predictions and electrical conductivity.

Rather than being attributable to a single modeling deficiency, the error probably arises  from the combined influence of the EOS through its impact on the predicted temperature and the conductivity model, emphasizing the strong coupling between thermodynamics and electrical transport in fully self-consistent simulations. 

These slight discrepancies aside, the simulations are in good agreement with the experimental data, thus providing a reliable basis for the design and optimization of future experiments.

\section{Conclusion\protect}

The objective of this work was to develop a comprehensive numerical framework capable of simulating a confined exploding foil experiment induced by pulsed-power.
 
To this end, we introduced a modeling approach for the electrical response of pulsed-power systems based on a simple yet robust current solver that incorporates an appropriate description of switch dynamics. This electrical model was validated through short-circuit current measurements performed on several pulsed-power generators, covering hundred of kiloamperes to megaamperes current amplitudes and microsecond timescales.

The thermodynamic evolution of the foil was simulated using a one-dimensional hydrodynamic code. As a first step, the energy deposition was imposed using experimentally measured electrical power, allowing direct comparison with experimental quantities. The electrical model was then coupled to the hydrodynamic solver through an experimentally derived electrical conductivity, yielding very good agreement with the measurements. 

In order to remove any dependence on experimental input signals, a self-consistent numerical approach was implemented using an electrical conductivity model. This coupled framework enables the self-consistent description of the complete experiment,  encompassing both the electrical response of the pulsed-power driver and the coupled electrical and hydrodynamic evolution of the load. The resulting simulations provide good agreement with the experimental data and effectively bracket the experimental behavior.

Beyond reproducing experimental quantities, the simulations also grant access to other properties that remain inaccessible to direct measurements. Moreover, it provides a straightforward methodology for testing different electrical conductivity and equation-of-state models under relevant conditions. Therefore, this numerical approach constitutes a powerful tool for interpreting existing experiments, guiding future designs and advancing the modeling of equation-of-state and transport properties.


\label{sec:conclusion}


\bibliographystyle{unsrt}
\bibliography{biblio_clean.bib}
\end{document}